\documentclass[manuscript,screen,nonacm]{acmart}
\acmConference[]{} 

\newcommand{\workshopname}{GenAICHI: CHI 2024 Workshop on Generative AI and HCI}
\newcommand{\licensedetails}{Licensed under a Creative Commons Attribution 4.0 International License (CC BY 4.0). Copyright remains with the author(s).}
\newcommand\extrafootertext[1]{
    \bgroup
    \renewcommand\thefootnote{\fnsymbol{footnote}}%
    \renewcommand\thempfootnote{\fnsymbol{mpfootnote}}%
    \footnotetext[0]{#1}%
    \egroup
}

\AtBeginDocument{ 
    \fancypagestyle{firstpagestyle}{
        \fancyhf{}
        \fancyfoot[L]{\sffamily\footnotesize \workshopname}%
        \fancyfoot[C]{\sffamily\footnotesize \thepage}
    }
    \fancyhf{}
    \fancyhead[L]{\sffamily\footnotesize\shorttitle}
    \fancyhead[R]{\sffamily\footnotesize\shortauthors}
    \fancyfoot[L]{\sffamily\footnotesize\workshopname}%
    \fancyfoot[C]{\sffamily\footnotesize\thepage}
    \extrafootertext{\licensedetails}
}
\usepackage{subcaption}

\begin{document}

\title{Case Study of GAI for Generating Novel Images for Real-World Embroidery}

\author{Kate Glazko}
\authornote{Authors contributed equally to the paper}
\email{glazko@uw.edu}
\orcid{0000-0002-7728-4764}
\affiliation{%
  \institution{University of Washington}
  \city{Seattle}
  \country{USA}
}

\author{Anika Arugunta}
\authornotemark[1]
\email{arugani9@cs.washington.edu}
\affiliation{%
  \institution{University of Washington}
  \city{Seattle}
  \country{USA}
}

\author{Janelle Chan}
\authornotemark[1]
\email{jc383@cs.washington.edu}
\affiliation{%
  \institution{University of Washington}
  \city{Seattle}
  \country{USA}
}

\author{Nancy Jimenez-Garcia}
\authornotemark[1]
\email{nancyjg8@cs.washington.edu}
\affiliation{%
  \institution{University of Washington}
  \city{Seattle}
  \country{USA}
}

\author{Tashfia Sharmin}
\authornotemark[1]
\email{tashfia@cs.washington.edu}
\affiliation{%
  \institution{University of Washington}
  \city{Seattle}
  \country{USA}
}

\author{Jennifer Mankoff}
\authornotemark[1]
\email{jmankoff@uw.edu}
\affiliation{%
  \institution{University of Washington}
  \city{Seattle}
  \country{USA}
}

\renewcommand{\shortauthors}{Glazko et al.}

\begin{abstract}
   In this paper, we present a case study exploring the potential use of Generative Artificial Intelligence (GAI) to address the real-world need of making the design of embroiderable art patterns more accessible. Through an auto-ethnographic case study by a disabled-led team, we examine the application of GAI as an assistive technology in generating embroidery patterns, addressing the complexity involved in designing culturally-relevant patterns as well as those that meet specific needs regarding detail and color. We detail the iterative process of prompt engineering custom  GPTs tailored for producing specific visual outputs, emphasizing the nuances of achieving desirable results that align with real-world embroidery requirements. Our findings underscore the mixed outcomes of employing GAI for producing embroiderable images, from facilitating creativity and inclusion to navigating the unpredictability of AI-generated designs. Future work aims to refine GAI tools we explored for generating embroiderable images to make them more performant and accessible, with the goal of fostering more inclusion in the domains of creativity and making. 
\end{abstract}

\maketitle

\section{Introduction and Background}
\label{sec:intro}
The growing presence of ChatGPT, DALLE-3, Midjourney, and other Generative Artificial Intelligence (GAI) tools has sparked discussions and research regarding their use in creativity, \cite{he2023wordart, vinchon2023creative}, artwork \cite{ko2023large, haase2023art, garciablending}, and making \cite{makatura2023can, liu20233dall, gao2022get3d, sonmez2021computer}. Emerging research has even highlighted the potential of using GAI to increase accessibility in these domains, such as easing the process of creating concept sketches for novel art or simplifying the process of learning about visual designs detailed in the research of Glazko \textit{et. al} \cite{glazko2023autoethnographic}. Utilizing GAI to help make creative work more accessible to people with disabilities could help address a critically un-met need. People with disabilities have historically been marginalized and left out of opportunities to participate in creative work \cite{holloway2021artists, ma2022ecological}, making/fabrication \cite{hurst2013making, leduc2013evaluating, steele2018accessible, seo2021scaffolding}, and even cultural heritage due to inaccessibility \cite{arenghi2017cultural, leahy2022barriers, lauria2016accessibility, fisher2023out}. 

Embroidery, an art form with deep cultural value and relevance across global communities, is one example of a creative domain where inaccessibility impacts the participation of disabled makers. Prior work has demonstrated how integrating embroidery with technology, such as using computionally-generated patterns, improves accessibility (e.g., \cite{cao2022computationally}). Likewise, research such as Zhang \textit{et al.} has explored the potential of digital tools to increase participation in culturally-relevant and traditional embroidery while facilitating learning in novices \cite{zhang2022mastersu}, while Ilieva \textit{et al.} have called out the potential for embroidery digitization to serve as a catalyst for sustainable practices in the fashion industry while preserving digital heritage \cite{ilievaapplication}. However, the process of creating novel designs, such as those suitable for machine embroidery, can itself be inaccessible to people with disabilities affecting motor coordination, dexterity, or cognition \cite{glazkogaidesign23}. State-of-the art tools for embroidery digitization such as Hatch \cite{hatchembroidery} which simplify the digitization part of the process do not simplify or make accessible the process of creating artwork suitable for embroidery. GAI has been proposed as one method for simplifying image generation and making the creative process more accessible \cite{glazko2023autoethnographic, glazkogaidesign23}.

In this work, we strive to understand how GAI can be used to foster both creativity and inclusion in the process of creating embroiderable artwork patterns. Designing artwork suitable for embroidery can be a complex process, with variables such as levels of detail and color having noted impacts on the quality of a digitization \cite{sewandgrew2023tips}. This work puts forward a case study of using GAI as an assistive technology to increase the accessibility of creating embroiderable images, conducted by disabled-led team of researchers with varying access needs. We present an auto-ethnographic case study where we use GAI as an access-enabling technology for embroidery and document our experiences. We finish by detailing important questions for our future work. By investigating the capabilities and limitations of LLMs in creative applications such as making computerized embroidery more accessible, we move closer to realizing a future where AI can contribute meaningfully to creative endeavors, fostering a more inclusive and diverse cultural landscape.

\section {Case Studies of GAI for Producing Embroidery Images}

\subsection{Method}
We conducted an initial evaluation of DALL-E 2 for embroidery (Fall '23) and a structured, multi-phase case study of incrementally improving DALL-E 3 with few-shot learning \cite{ahmed2022few} (Winter '24). The initial evaluation was performed by one author (A1) with disabilities, and the follow-up case study was conducted by a team of five researchers (A1-A5) with varying disabled identities/access needs, and with a variety of intersectional identities (i.e. racial identities, genders, LGBTQIA+ status). For both studies, DALL-E was used to generate images (PNGs) that were converted to vectors through Adobe software. We note that using PNG to SVG conversion introduced complexity and impacted accessibility in these early explorations, but GPT-4's capability for producing SVG outputs was not performant (as shown in Fig. \ref{fig:svg}). In future prototypes of a LLM-based embroidery tool, we can automate this process through libraries without creating any additional user interaction. Embrilliance and Inkscape were used interchangeably to convert image vectors to embroidery files (.JEF or .PES). Two computerized embroidery machines, a Brother SE1900 and Janome Skyline 9 were used in our evaluations, with no notable differences in performance due hardware capabilities.

\subsection{Initial Exploration}
\begin{figure}
    \centering
    \begin{subfigure}[t]{0.3\textwidth}
        \centering
        \includegraphics[height=1.3in]{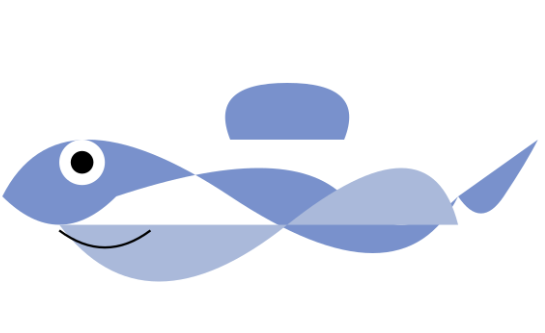}
        \caption{Best example of generated SVG dolphin (not recognizable)}
        \Description{Abstract, blocky SVG “dolphin” made of overlapping shapes; not recognizable, looks more like an abstract art of a whale. This shows poor direct SVG generation for embroidery.}
        \label{fig:svg}
    \end{subfigure}
    \hfill
    \begin{subfigure}[t]{0.3\textwidth}
        \centering
        \includegraphics[height=1.3in]{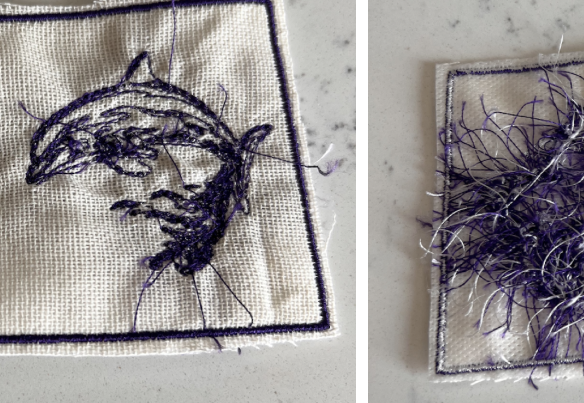}
        \caption{Generated PNG dolphin with no real-world prompt constraints (led to embroidery errors)}
        \Description{Photo of a hoop stitch-out of an unconstrained PNG dolphin. The embroidery is intricate and good-quality, but too many small details cause jagged lines and failed embroidery on the reverse.}
        \label{fig:dolphin}
    \end{subfigure}%
    \hfill
    \begin{subfigure}[t]{0.3\textwidth}
        \centering
        \includegraphics[height=1.3in]{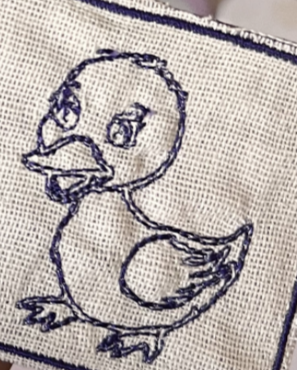}
        \caption{Generated PNG duck with real-world prompt instructions (i.e. image complexity, line width, position)}
        \Description{Clean, centered line drawing of a duck with minimal detail and adequate line width suited to embroidery after extensive prompting.}
        \label{fig:teaser:duck}
    \end{subfigure}%
    \caption{DALL-E 2 Initial Exploration}
    \label{fig:initialexplore}
\end{figure}

An initial exploration into using GAI for creating machine-embroiderable images was conducted in Fall 2023 by A1, a disabled researcher whose cognitive and motor disabilities  negatively impact their ability to physically or digitally draw designs. A1 wanted to see if DALL-E 2 could be used to improve access for themselves and produce novel designs for embroidery on-demand without the need to draw.  A1 recorded a detailed log of their experiences using DALL-E 2 to produce the images that could translate to real-world embroidered patterns. Issues encountered in generating images through DALL-E 2 included: images with properties that resulted in failed embroidery runs: images with too many small details as shown in \ref{fig:dolphin}, images with jagged lines, images that are cut off, images that are not centered, amonst other issues. A1 had to constantly update their prompt to explicitly define properties needed to produce an image that was possible to embroider such as the one seen in Fig. \ref{fig:teaser:duck}, a repetitive and time-consuming process.

\subsection{GAI and Embroiderable Images Case Study}
The initial exploration demonstrated the need for constant prompt engineering to produce satisfactory results from DALL-E 2. The team wanted to see if there were ways to reduce user burden and embed a set of real-world image properties into GAI \textit{by default}. To do this, we (A2-A5) iteratively attempted to produce embroiderable images with specific properties embedded in GPT-4 through the use of the no-code, custom GPTs tool (CU-GPTs). CU-GPTs improves performance for specific through modifying directive through written tasks and through the inclusion of training files and images. A2-A5 each created a CU-GPTs and assessed its performance in producing specific visual outputs with real-world properties over the course of two months. A1 interacted with and evaluated the CU-GPTs as an end-user would with no prompt engineering, producing machine embroidery files from the generated SVGs and fabricating the designs. Few-shot learning approaches and optimizations were documented when they occurred. We each independently wrote down our experiences with CU-GPTs in a shared document. The output CU-GPTs were evaluated qualitatively on a weekly basis, and from reviewing our logs, we identified common themes.

\subsubsection{Optimizing GAI for Producing Specific Image Outputs} 
\begin{figure}
    \centering
    \begin{subfigure}[t]{0.2\textwidth}
        \centering
        \includegraphics[height=1.25in]{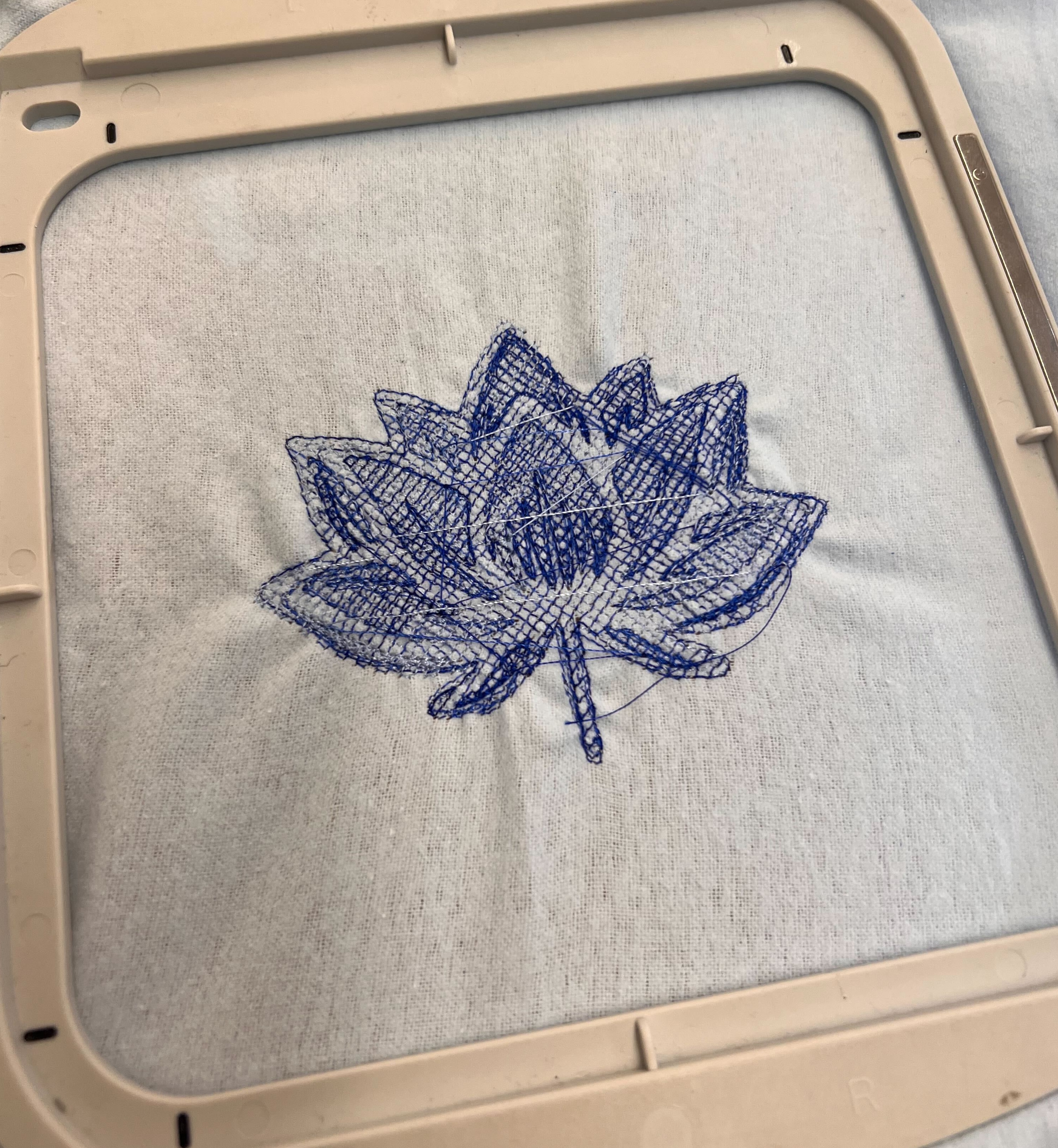}
        \caption{\textbf{Success:} Machine embroidered generated lotus image}
        \Description{Completed lotus motif stitched in an embroidery hoop; low-detail, yet aesthetic design demonstrating successful GAI-to-embroidery translation.}
        \label{fig:lotus}
    \end{subfigure}
    \hfill
    \begin{subfigure}[t]{0.2\textwidth}
        \centering
        \includegraphics[height=1.25in]{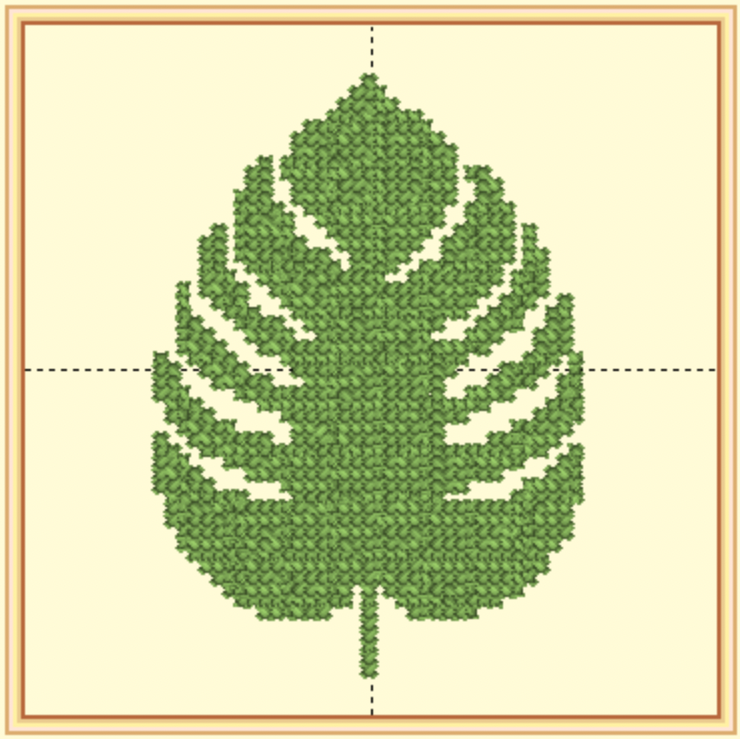}
        \caption{\textbf{Success:} Generated Monstera converted to an Embrilliance embroidery file}
        \Description{Embrilliance screenshot of a Monstera leaf with manageable paths and fills—example of a customer GPT output ready for embroidery.}
        \label{fig:monstera}
    \end{subfigure}%
    \hfill
    \begin{subfigure}[t]{0.2\textwidth}
        \centering
        \includegraphics[height=1.25in]{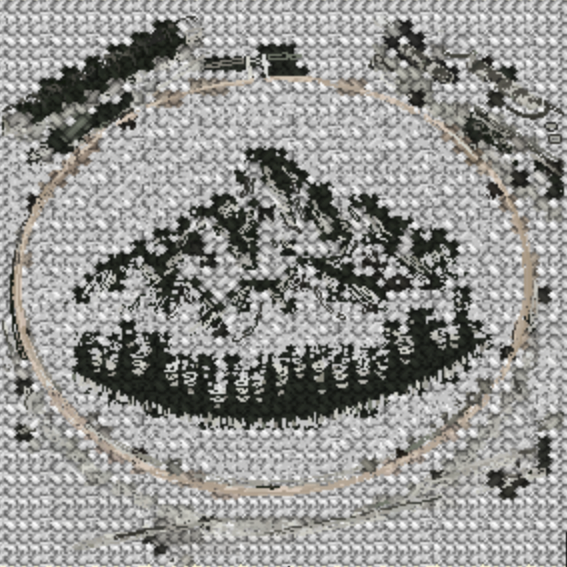}
        \caption{\textbf{Error:} Generated Mountain Scene with unwanted background resulting in 500+ SVG paths}
        \Description{SVG of a Mountain landscape with heavy background texture; excessive (500+) SVG paths make the design impractical to embroider.}
        \label{fig:background}
    \end{subfigure}%
    \hfill
    \begin{subfigure}[t]{0.2\textwidth}
        \centering
        \includegraphics[height=1.25in]{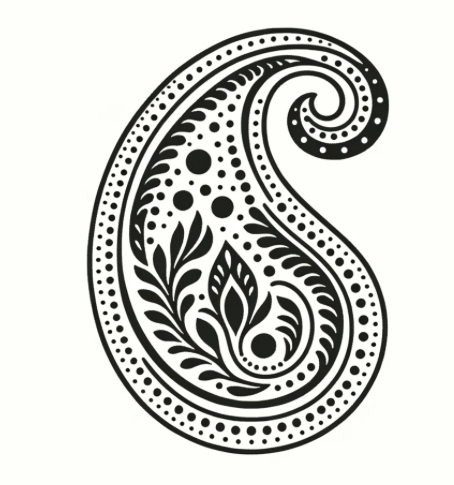}
        \caption{\textbf{Error:} Generated Paisley with excessive dots, failed at SVG conversion stage}
        \Description{A highly ornate paisley filled with many dots. The overly-complex detail leads to SVG conversion failure despite constraints given in prompting.}
        \label{fig:paisley}
    \end{subfigure}%
    \caption{Custom GPTs Few-Shot Learning Case Study}
    \label{fig:on-customgpts}
\end{figure}

Each member of our team produced a CU-GPT with unique physical properties and constraints. The types of embroiderable content the CU-GPTs were trained to produce included images of: paisleys, lotuses, monstera leaves, and mountain scenery. The CU-GPTs were trained with a no-code, few-shot learning approach and were iteratively improved through prompt engineering and adding training files. 
The additional training files and iterations improved the performance of some of the CU-GPTs, which more consistently began to produce images that obeyed specific, real-world properties. However, we noted that very explicit instructions, such as specifically banning CU-GPTs from images that had embroidery textures built-in (with texture resulting in overly-complex SVGs), were necessary for improved performance.
\begin{quote}
 Final Prompt: Generate an image of a simple, minimal-detail Monstera leaf silhouette with a clean outline. Strictly avoid any appearance of embroidery. Ensure that the image is centered and that it is only the outline of the Monstera leaf with minimal fenestrations. The image should have a white background and should not include small/intricate details. It is difficult for my embroidery machine to print out small details, so please avoid them. (A3, Monstera CU-GPTs)
\end{quote}
A5 likewise found specific, detailed instruction was needed in order to produce a clean, plain background for embroiderable images of mountain scenery. In one instance, she received an image she was very satisfied with, but CU-GPTs rendered the image inside of a visual embroidery hoop, resulting in a pattern too complex to embroider as shown in Fig. \ref{fig:background}. 

\subsubsection{Not Fixed By Few-Shot: Instructions Ignored and Inaccuracies}
A surprising frustration of using GAI was illustrated by A4, who tried to use a CU-GPTs to produce images of paisleys, describing her initial experience with CU-GPTs "I also liked how it was able to emulate cultural patterns inside the paisley" (A4). However, A4 noted how the paisleys produced by the CU-GPTs consistently produced intricate, complicated details that would be difficult to embroider, specifically dots. A4 made multiple attempts to modify the CU-GPTs to exclude dots from the images through modifying the directive and uploading sample file images without any dots featured. 
A4's CU-GPTs was occasionally able to identify that the resulting image had dots, and would issue an apology and re-generate the image. Most of the time, the resulting image would have no fewer dots. We found it surprising that despite providing sample images with paisleys and no dots, CU-GPTs was not able to produce outputs without dots. A Google Image search yieled multiple examples of paisleys without dots. 

\subsubsection{CU-GPTs (Mostly) Translated to the Real-World}
A1 interacted with the CU-GPTs as they previously had with out-of-the-box DALL-E 2 and GPT-4. They converted the generated images to SVGs and imported them to Embrilliance, counting the number of paths, jumps, and difficult-to-embroider details. Overall, A1 described encountering fewer issues with the CU-GPTs images when compared to the initial evaluations; encountering much fewer issues with images being off-center, seeing fewer SVGs with excessive paths, and requiring fewer back-and-forth conversations with GPT. A1 was more satisfied overall with the embroidery files due to the ease of creating successful designs such as the Monstera shown in Fig. \ref{fig:monstera}, produced from the CU-GPTs images than in the initial exploration, describing the lotus they successfully embroidered as "beautiful", as shown in Fig.
\ref{fig:lotus}, describing "It is easier to check if an image will successfully embroider than trying to draw something like a lotus by hand with a stylus" (A1).

\section{Conclusion, Future Work, and Open Questions for Workshop}
This study highlights the benefits and challenges of using generative AI, specifically custom GPTs (CU-GPTs), in creating images suitable for embroidery with real-world constraints. Our future work will focus on combining instructions from the most consistently-performant CU-GPTs such as the one seen in Fig.\ref{fig:monstera} and Fig. \ref{fig:lotus} to create an optimal prompt instruction for a CU-GPTs capable of producing more universal imagery. We will quantitatively evaluate our CU-GPTs and assess whether a more complex approach, such as using images generated from our CU-GPTs, to train our own model specializing in images for embroidery is needed. Lastly, we will design and test an accessible, GAI-based embroidery generation tool with disabled creators and makers.

Through our work, we hope to answer the question: "Do generative AI algorithms contribute needed serendipity to the design process—or simply randomness—or worse, chaos?" We believe our work has demonstrated a mix of both-- the potential to bring serendipity and joy of making to individuals who previously lacked access, coupled with the boundless frustrations of attempting to control and manipulate the performance of a black-box entity.

{\small
\bibliographystyle{GAIEmbroidery}
\bibliography{GAIEmbroidery}
}

\end{document}